\documentstyle[11pt]{article} 
\input epsf

 \newcommand{\be}{\begin{equation}}
 \newcommand{\ee}{\end{equation}}
\newcommand{\bea}{\begin{eqnarray}} \newcommand{\eea}{\end{eqnarray}}
\begin{document}

\begin{center}
{\LARGE{\bf Non-hermitian Exact Local Bosonic Algorithm for Dynamical Quarks}}\\[1cm]

\vspace*{1cm}
{\Large A. Borrelli$^a$, Ph. de Forcrand$^{b,}$\footnote{forcrand@scsc.ethz.ch},
A. Galli$^{c,}$\footnote{galli@mppmu.mpg.de}}\\[0.3cm]
$^a${\small \em Universita di Roma "La Sapienza", I-00198 Roma, Italy}\\
$^b${\small \em SCSC, ETH-Zentrum, CH-8092 Z\"urich, Switzerland}\\
$^c${\small \em Max-Planck-Institut f\"ur Physik, D-80805 Munich, Germany}
\vspace*{1cm}
\end{center}
\begin{abstract}
We present an exact version of the local bosonic algorithm for the simulation of
dynamical quarks in lattice QCD. This version is based on a non-hermitian 
polynomial approximation of the inverse of the quark matrix. 
A Metropolis test corrects the systematic errors. Two variants of this test are 
presented. For both of them, a formal proof is given that this Monte Carlo 
algorithm converges to the right distribution. 
Simulation data are presented for different 
lattice parameters. The dynamics of the algorithm and 
its scaling in lattice volume and quark mass are investigated.
\end{abstract}
\newpage

\section{Introduction}
As the accuracy of quenched QCD simulations improves, the systematic error
caused by the quenched approximation becomes more and more important.
Accurate full QCD simulations are needed, for which hardware performance
improvement appears insufficient. Theoretical and algorithmic progress
is necessary.\\
 
The search for more efficient full QCD algorithms has motivated substantial
activity, both within the classical Hybrid Monte Carlo (HMC) and in
the alternative method of the local bosonic algorithm.
This last algorithm was proposed by L\"uscher in a hermitian version 
\cite{luescher1}.
The idea was to approximate the full QCD partition function using a local
bosonic action based on a polynomial approximation of the inverse of the
squared Wilson fermion matrix. The aim was to obtain an algorithm which
does not need the explicit inversion of the fermion matrix and which
only uses local, finite-step-size updates, contrary to HMC.
If one evaluates the low-lying spectrum of the matrix, the systematic
errors due to the approximation can be corrected, either in the measurement
of observables \cite{luescher1} or
directly by a Metropolis test \cite{peardon}.
Unfortunately, it turned out that the autocorrelation time was very
large \cite{luescher2, beat1} and the correction of the systematic errors 
using the spectrum very costly \cite{borrelli}. \\
 
Recently, several significant improvements have been proposed to improve the
efficiency of this algorithm: (a) an inexpensive stochastic Metropolis test
\cite{forcrand1}; (b) a non-hermitian polynomial approximation \cite{forcrand1};
(c) a simple even-odd preconditioning \cite{beat2}; (d) a two-step approximation
for non-even number of flavors \cite{montvay}. Preliminary studies combining
(a) and (b) looked very promising \cite{forcrand2}. Here we combine (a), (b) 
and (c) for two-flavor QCD, and we present a detailed description of the 
algorithm and a formal proof of its correctness. We illustrate the 
effectiveness of our algorithm with representative Monte Carlo results. We 
successfully model its static properties (the Monte Carlo acceptance), and try 
to disentangle its dynamics, using both Monte Carlo measurements and a 
toy-model.\\

From our analysis it appears that this version of the local bosonic algorithm 
is a real alternative to the classic HMC.
Its scaling in lattice volume and quark mass compares favorably
with respect to HMC. Thus our algorithm becomes increasingly attractive
at low quark mass on large lattices. 
In addition, because it uses local updating techniques,
it is not affected by the accumulation of roundoff errors which can marr
the reversibility of HMC in such cases.

\newpage
\section{Description of the algorithm}

\subsection{The non-hermitian exact local bosonic algorithm}

The full QCD partition function with two fermion flavors is given by
\be Z=\int [dU] |\det D|^2 e^{-S_G[U]}
\label{fullQCD}
\ee where $D$ represents the fermion matrix and $S_G$ denotes the pure gauge action. This partition function can be approximated introducing a polynomial
approximation of the inverse of the fermion matrix.  \be |\det
D|^2=\det D^\dagger\det D\simeq \frac{1}{|\det P(D)|^2}
\label{app}
\ee The polynomial $P(z)=\prod_{k=1}^n(z-z_k)$ of degree $n$ is
defined in the complex plane and approximates the inverse of $z$. 
The approximation is controlled by the
degree $n$ of the polynomial and the position of its roots in the
complex plane. A complete discussion of the choice of the roots is
given in \cite{forcrand1}.  
In our work we have chosen the roots on an ellipse centered in $c$ and with focal distance $f$
\be
z_k=c(1-\cos\theta_k)-i\sqrt{c^2-f^2}\sin\theta_k\label{rootdef}
\ee
where $\theta_k=\frac{2\pi k}{n+1}$. For our simulation at $\beta=0$ 
we have chosen a circle
with $c=1$ and $f=0$.
Since we are investigating full QCD with two flavors, the determinant 
$|\det P(D)|^2$ manifestly factorizes into positive pairs, for {\em any}
choice of the $z_k$'s (real or complex) and {\em any} degree $n$ (even or odd):
\be |\det
P(D)|^2=\mbox{constant}\times\prod_{k=1}^{n}\det(D-z_k)^\dagger(D-z_k) \ee 

The
$\frac{1}{|\det P(D)|^2}$ 
term of the approximation (\ref{app}) can be expressed
by a Gaussian integral over a set of boson fields $\phi_k$ ($k=1,...,n$)
with color and Dirac indices and with partition function 
\be
Z_b=\int\prod_{k=1}^n [d\phi_k][d\phi^\dagger_k]
e^{-\phi^\dagger_k(D-z_k)^\dagger(D-z_k)\phi_k}
\label{zb}
\ee
The full
QCD partition function (\ref{fullQCD}) is then approximated by \be
Z\simeq \int [dU] Z_b[U] e^{-S_G[U]}
\label{lus}
\ee Making use of the locality of the L\"uscher action
\be
S_L=S_G+S_b
\ee
where 
\be 
S_b=\sum_{k=1}^nS_b^k=\sum_{k=1}^n 
\phi^\dagger_k(D-z_k)^\dagger(D-z_k)\phi_k\label{locac}
\ee
we may now simulate the partition function (\ref{lus}) by locally updating
the boson fields and the gauge fields, using heat-bath and
over-relaxation algorithms. \\

We consider the update $(U,\phi)\rightarrow (U',\phi')$ of the gauge and boson 
fields consisting of a "trajectory" or sequence of $m$ local updates according 
to the approximate partition function (\ref{lus}). 
We choose an update sequence of these fields which satisfies detailed balance with respect to the L\"uscher action $S_L=S_G+S_b$. This means that the transition probability from an old  configuration $(U,\phi)$ to a new one $(U',\phi')$ satisfies
\be
\frac{P_{(U,\phi)\rightarrow(U',\phi')}^{L}}{
 P_{(U',\phi')\rightarrow(U,\phi)}^{L}}=e^{-(S_L(U',\phi')-S_L(U,\phi))}\label{dbl}
\ee
 This is achieved by symmetrically ordering the update steps of the $U$ and 
$\phi$ fields (over-relaxation and heat-bath for the boson fields; 
over-relaxation for the gauge fields) so that they be invariant under reversal
of the trajectory. Additionally, one can organize the updates on random 
sublattices. Different organizations are also possible.
In our simulations we have tested different combinations of the updating steps. 
A quite efficient one is composed of one heat-bath of the boson fields, 
$m$ alternated over-relaxations of the boson and the gauge fields, and
a further over-relaxation and heat-bath of the boson fields to symmetrize 
the trajectory.\\

The simulation of full QCD can be obtained from (\ref{lus}) by correcting
the errors due to the approximation through a Metropolis test at the end of each trajectory.
Introducing the error term in the partition function we obtain 
\bea 
Z&=&\int [dU]|\det
D|^2\frac{|\det P(D)|^2}{|\det P(D)|^2}e^{-S_G[U]}=
\int [dU] |\det(DP(D))|^2 Z_b[U] e^{-S_G[U]}=\nonumber\\
&=&\int [dU] [d\phi] [d\phi^\dagger]|\det(DP(D))|^2e^{-S_L(U,\phi)}\label{exact}
\eea
where $\phi$ represents the sets of all boson field families.
The correction term $ |\det(DP(D))|^2$ can be evaluated in two different 
ways.\\

The first one consists in estimating the average determinant $ <|\det(DP(D))|^2>$
using a noisy estimator of the Gaussian integral
\be
\int [d\eta][d\eta^\dagger]e^{-|[DP(D)]^{-1}\eta|^2}\label{etaint}
\ee
The strategy of this method is to update the $(U,\phi)$ fields such that the probability of finding a particular configuration is proportional to $e^{-S_L}$ and then perform a Metropolis test defining an acceptance probability $P^A$ in 
terms of the noisy 
estimation of the correction $|\det(DP(D))|^2$. 
In this case the full transition probability of the algorithm $P=P^LP^AP^{HB}$
satisfies detailed balance 
\be
\frac{<P_{(U,\phi)\rightarrow (U',\phi')}>}{
      <P_{(U',\phi')\rightarrow (U,\phi)}>}=
\frac{ |\det(D'P(D'))|^2e^{-S_L(U',\phi')}}{
 |\det(DP(D))|^2e^{-S_L(U,\phi)}}\label{33}
\ee
{\em after} averaging over the $\eta$ fields which are generated by a Gaussian heatbath according to the probability density $P^{HB}$.
Here we call $D'$ and $D$ the fermion matrix for the new and old gauge 
configurations respectively.\\

The second method consists in expressing the correction term $ |\det(DP(D))|^2$
directly by a Gaussian integral and incorporating the dynamics of the correction field $\eta$ in the partition sum (\ref{exact})
\bea
Z&=&\int [dU][d\eta][d\eta^\dagger][d\phi] [d\phi^\dagger] e^{-S_L[U]}
e^{-|[DP(D)]^{-1}\eta|^2}=\nonumber\\
&=&\int [dU][d\eta][d\eta^\dagger]
 [d\phi] [d\phi^\dagger]e^{-S_{exact}(U,\phi,\eta)}
\label{exact2}
\eea
by defining a new exact action
\be
S_{exact}(U,\phi,\eta)=S_{L}(U,\phi)+S_{C}(U,\eta)\label{nonoisy}
\ee
where 
\be
S_{C}(U,\eta)=|[DP(D)]^{-1}\eta|^2\label{sc}
\ee
is the correction action. In carrying out the simulation one generates 
configurations of $(U,\phi)$ and $\eta$ such that the probability of 
finding a particular configuration is proportional to $\exp(-S_{exact})$. 
Also in this case the strategy is to alternatively update the $(U,\phi)$ 
fields and the $\eta$ fields. The Metropolis acceptance  probability 
$P^A$ is not defined using a noisy estimation of the correction 
$ |\det(DP(D))|^2$, but directly using the exact action (\ref{nonoisy}) 
so that the transition probability of the algorithm $P=P^LP^AP^{HB}$
satisfies detailed balance 
\be
\frac{P_{(U,\phi,\eta)\rightarrow (U',\phi',\eta')}}{
      P_{(U',\phi',\eta')\rightarrow (U,\phi,\eta)}}=
\frac{ e^{-S_{exact}(U',\phi',\eta')}}{
 e^{-S_{exact}(U,\phi,\eta)}}
\ee
{\em without} the need to average over $\eta$.
We have tested both methods and both seem equally efficient.

\subsubsection{Noisy Metropolis test}
The idea of the noisy estimator is to generate for each trajectory a field 
$\eta$ using a heatbath according to the Gaussian distribution
\be
P_{\eta}^{HB}(U)=
\frac{R}{|\det DP(D)|^2}e^{-|[DP(D)]^{-1}\eta|^2}
\label{noisypppp}
\ee
(where $R$ is a normalization constant),
and accept the new configuration 
$(U',\phi')$ with the acceptance probability


\bea
P^{A}_{(U,\phi)\rightarrow(U',\phi')}&&=
\min\left(1,e^{-|[D'P(D')]^{-1}\eta|^2+|[DP(D)]^{-1}\eta|^2}\right)
\nonumber\\
&&=\min\left(1,E^A_{U\rightarrow U'}\right)
\eea
The probability density responsible for the Metropolis correction is then 
the product
\be
P^{C}_{(U,\phi)\rightarrow(U',\phi')}=
P^{HB}_\eta(U)\,P^{A}_{(U,\phi)\rightarrow(U',\phi')}
\ee
For a single trajectory the correction term is estimated by one $\eta$ field only. Because a trajectory making a transition  
$(U,\phi)\rightarrow (U',\phi')$ can occur infinitely many times one can average over the $\eta$ field and using
\bea
&& <P^C_{(U,\phi)\rightarrow(U',\phi')}>=\\
&&=\frac{R}{|\det DP(D)|^2}\left[
\int_{E^A>1} [d\eta][d\eta^\dagger] e^{-|[DP(D)]^{-1}\eta|^2}+
\int_{E^A<1} [d\eta][d\eta^\dagger] e^{-|[D'P(D')]^{-1}\eta|^2}
\right]\nonumber
\eea
and 
\be
E^{A}_{U\rightarrow U'}=\frac{1}{
E^{A}_{U'\rightarrow U}}
\ee
one can see that detailed balance (\ref{33}) is satisfied. 
Because in the probability density (\ref{noisypppp}) we have the inverse of $DP(D)$ we can organize the Gaussian heatbath by generating a Gaussian spinor 
$\chi$ with variance one and obtaining the $\eta$ field by assigning
$\eta=DP(D)\chi$.  
It is then convenient to express the acceptance probability $P^A$ using this substitution 
\be
P^A_{(U,\phi)\rightarrow(U',\phi')}=
\min\left(1,e^{-|W\chi|^2+|\chi|^2}\right)
\label{32}
\ee
where $W=[D'P(D')]^{-1}DP(D)$ as suggested in \cite{forcrand1}. \\

\subsubsection{Non-noisy Metropolis test}

Let us consider now the updating of the $\eta$ field when its dynamics is 
incorporated in the partition sum by the exact action (\ref{nonoisy}). 
We want to update this field by a Gaussian heatbath.
The probability density $P^{HB}_{\eta\rightarrow\eta'}(U,U') $ for obtaining 
a field $\eta'$ at the end of a trajectory, starting from the field $\eta$ 
at the beginning of a trajectory, can be chosen trivially independent from 
$\eta$ but has to depend upon 
the gauge fields $U$ or $U'$. In addition it has to satisfy exactly 
detailed balance with respect to the  action $S_C$ when keeping 
the gauge fields $U$ and $U'$ fixed.
This is achieved by introducing an arbitrary ordering of the gauge configurations, for example
\be
U\succ U'\mbox{  if }S_G(U)>S_G(U')
\ee
and choosing the Gaussian heatbath probability density 
\be
P_{\eta\rightarrow\eta'}^{HB}(U,U')=
\left\{\begin{array}{lr}
\frac{R}{|\det DP(D)|^2}e^{-S_C(U,\eta')}&\mbox{ if } U\succeq U'\\
\frac{R}{|\det D'P(D')|^2}e^{-S_C(U',\eta')}&\mbox{ if } U\prec U'
\end{array}
\right.\label{pppp}
\ee
Here $R$ is a normalization constant and $S_C$ is defined in (\ref{sc}). Detailed balance is then satisfied for fixed gauge fields
\be
\frac{P_{\eta\rightarrow\eta'}^{HB}(U,U')}{P_{\eta'\rightarrow\eta}^{HB}(U',U)}=
\left\{\begin{array}{lr}
e^{-(S_C(U,\eta')-S_C(U,\eta))}&\mbox{ if } U\succeq U'\\
e^{-(S_C(U',\eta')-S_C(U',\eta))}&\mbox{ if } U\prec U'
\end{array}
\right.
\label{seta1}
\ee
The crucial point in this method is that
the arbitrary ordering of the gauge fields and the choice 
(\ref{pppp}) ensure that in the ratio (\ref{seta1}) the unwanted determinants cancel. The cancellation of the determinants allows us to construct a non-noisy algorithm. 
Because in the action $S_C$ we have the inverse of $DP(D)$ (or $D'P(D')$) 
the Gaussian heatbath is easily organized as follows: one generates a Gaussian 
spinor $\chi$ with variance one and obtains the $\eta'$ field by assigning
\bea
\eta'&=&
\left\{\begin{array}{lr}
DP(D)\chi&\mbox{ if } U\succeq U'\\
D'P(D')\chi&\mbox{ if } U\prec U'
\end{array}
\right.
\label{assig}
\eea
For each trajectory a Gaussian
spinor $\chi$ is chosen.
At the end of each trajectory a Metropolis test is performed to be able to correct the approximate update of the $U$ and $\phi$ fields.  
In order to ensure that detailed balance is satisfied with respect to the 
exact action $S_{exact}$  we accept the proposed change in the $U$ and $\phi$ 
fields with an acceptance probability 
$P^A_{(U,\phi,\eta)\rightarrow(U',\phi',\eta')}$ 
which is chosen such that
\be
\left.
\frac{
P_{(U,\phi)\rightarrow(U',\phi')}^{L}
P_{\eta\rightarrow\eta'}^{HB}(U,U')
P^A_{(U,\phi,\eta)\rightarrow(U',\phi',\eta')}
}{
P_{(U',\phi')\rightarrow(U,\phi)}^{L}
P_{\eta'\rightarrow\eta}^{HB}(U',U)
P^A_{(U',\phi',\eta')\rightarrow(U,\phi,\eta)}}
\right|_{\chi}
=e^{-(S_{exact}(U',\phi',\eta')-S_{exact}(U,\phi,\eta))}\label{db}
\ee 
For every choice of Gaussian spinor $\chi$.
We can choose the probability $P^{A}$ dependent on $\eta$ and  $\eta'$ according to 
\be
P^A_{(U,\phi,\eta)\rightarrow(U',\phi',\eta')}=
\left\{\begin{array}{lr}
\min\left(1,e^{-S_C(U',\eta')+S_C(U,\eta')}\right)
&\mbox{ if } U\succeq U'\\
\min\left(1,e^{-S_C(U',\eta)+S_C(U,\eta)}\right)
&\mbox{ if } U\prec U'
\end{array}
\right.
\label{pacc}
\ee
For every choice of Gaussian
spinor $\chi$ it is clear that the probability $P^A$ (\ref{pacc}) satisfies
\be
\frac{
P^A_{(U,\phi,\eta)\rightarrow(U',\phi',\eta')}
}{
P^A_{(U',\phi',\eta')\rightarrow(U,\phi,\eta)}}
=
\left\{\begin{array}{lr}
e^{-(S_C(U',\eta')-S_C(U,\eta'))}
&\mbox{ if } U\succeq U'\\
e^{-(S_C(U',\eta)-S_C(U,\eta))}
&\mbox{ if } U\prec U'
\end{array}
\right.
\ee
and using eq. (\ref{dbl}) and (\ref{seta1}) we prove that 
detailed balance (\ref{db}) holds. \\

It is convenient to express the acceptance probability (\ref{pacc}) using (\ref{assig})
\be
P^A_{(U,\phi,\eta)\rightarrow(U',\phi',\eta')}=
\left\{\begin{array}{lr}
\min\left(1,e^{-|W\chi|^2+|\chi|^2}\right)&\mbox{ if } U\succeq U'\\
\min\left(1,e^{+|W^{-1}\chi|^2-|\chi|^2}\right)&\mbox{ if } U\prec U'
\end{array}
\right.\label{df}
\ee
where $W=[D'P(D')]^{-1}DP(D)$ as suggested in \cite{forcrand2}. \\

\subsubsection{Summary}

The algorithm can be summarized as follows:
\begin{itemize}
\item Generate a new Gaussian spinor $\chi$ with variance one.
\item Update locally the boson and gauge fields $m$ times in a reversible way.
\item Accept/reject the new configuration according to the Metropolis acceptance
probability (\ref{32}) for the noisy version 
or (\ref{df}) for the non-noisy version. 
\end{itemize}

In order to evaluate the Metropolis acceptance probability we have to solve 
linear system involving $DP(D)$ or $D'P(D')$, 
for which we use the BiCGstab algorithm
 \cite{schilling}. This linear system is very well conditioned because $P(D')$ 
(or $P(D)$) approximates the inverse of $D'$ (or $D$). 
The cost for solving it is moderate 
and scales like the local updating algorithms in the volume and quark mass. \\

We emphasize that the algorithm remains exact for any choice of the polynomial
$P$. The only requirement is that its roots $z_k$ come in complex conjugate
pairs or be real [Note that for $n_f$ even, real roots and an odd degree $n$
are possible]. If the polynomial approximates the inverse of the fermion matrix 
well, the acceptance of the Metropolis correction will be high; 
if not the acceptance will be low. 
The number and location of the roots $z_k$ in the complex plane determine the 
quality of the approximation and hence the acceptance. 
Since the algorithm is exact for all polynomial, we do not need any a priory 
knowledge about the spectrum of the fermion matrix.

\subsection{Local updating algorithms}
We discuss in this section how we have organized the local updating algorithm 
for the boson and gauge fields. For the updating of these fields we use 
standard heatbath and over-relaxation algorithms.
\subsubsection{Boson field update}
We have a set of $n$ families of boson fields $\phi=\{\phi_k\}$, each one 
associated with a root of the polynomial. Different families do not interact 
with each other. For a given family $k$, the boson field $\phi_k$ couples only 
to the gauge fields. The bosonic action $S_b^k$ has a next-to-nearest neighbor 
interaction term which couples a field $\phi_k(x)$ at point $x$ with all its 
neighbors $\phi_k(y)$ at points $y$ within distance $|x-y|\leq\sqrt{2}$ 
(Euclidean-norm).   
The local updating algorithm can then be organized by visiting simultaneously 
all lattice points at a mutual distance larger than $\sqrt{2}$. 
Dividing the lattice in $2^4$-hypercubes and labeling their $8$ diagonals, all
points sitting on the diagonals with the same label have the wanted mutual distance. This organization is particularly convenient for 
a MIMD parallel machine, since it requires few synchronizations. 
For SIMD machines much simpler organizations are possible. 
The set of all these diagonals defines 
$8$ sublattices. These sublattices are visited randomly to ensure reversibility.
As a function of $\phi_k(x)$  when all the fields on different sublattices are kept fixed, the local boson action takes the form
\be
S_b^k=A_k|\phi_k(x)|^2+[b_k]^\dagger\phi_k(x)+[\phi_k(x)]^\dagger b_k+\mbox{constant}
\ee
Using the notation of the Wilson fermion matrix $D$ acting on a field $\phi$
at a point $x$ 
\be
[D\phi](x)=\phi(x)-K\sum_{\mu=0}^3[U(x,\mu)(1-\gamma_\mu)\phi(x+\hat \mu)
+U(x-\hat\mu,\mu)^\dagger(1+\gamma_\mu)\phi(x-\hat\mu)]\label{fermionmatrix}
\ee
we have after some simple algebra
$$
A_k=1+16K^2-2Re(z_k)+|z_k|^2
$$
and
$$
b_k=[D^\dagger D\phi_k](x)-\bar z_k[D\phi_k](x)-z_k[D^\dagger\phi_k](x)
-(1+16K^2-2Re(z_k))\phi_k(x)
$$
The heatbath is then simply given by
\be
\phi_k(x)\longleftarrow A_k^{-1/2}\chi-A_k^{-1}b_k
\ee
where $\chi$ is a Gaussian spinor with variance one. The over-relaxation is given by
\be
\phi_k(x)\longleftarrow -\phi_k(x)-2A_k^{-1}b_k
\ee
\subsubsection{Gauge field update}
The updating of the gauge fields can be organized by dividing the lattice into 
even and odd sites. This defines two sublattices. All gauge links $U(x,\mu)$ starting from a sublattice can be updated simultaneously for a fixed direction $\mu$.
When all other links are kept fixed the action takes the form
\be
S=Re\left(tr\{U(x,\mu)(F_U+F_\phi)^\dagger\}\right)+\mbox{constant}
\ee
where $F_U$ is the usual gauge link staple around $U(x,\mu)$ and $F_\phi$ is 
the staple given by the boson fields, which can be expressed by
the matrix $F_\phi=\sum_k v_k[w_k]^\dagger$ with
\bea
v_k&=&\left\{(1-\frac{K}{2}) -z_k -[(1+\frac{K}{2}) -
z_k ]\gamma_\mu\right\}\phi_k(x)-K (1-\gamma_\mu)B^k_\mu(x)\nonumber\\
w_k&=&\left\{(1-\frac{K}{2}) -z_k +[(1+\frac{K}{2}) -
z_k ]\gamma_\mu\right\}\phi_k(y)-K (1+\gamma_\mu)B^k_\mu(y)\nonumber
\eea
where $y=x+\hat\mu$ and
$$
B^k_\mu(z)=\sum_{\nu\neq\mu}\{U(z,\nu)(1-\gamma_\nu)\phi_k(z+\hat\nu)+
U(z-\hat\nu,\nu)^\dagger(1+\gamma_\nu)\phi_k(z-\hat\nu)\}
$$ 
Once the staple $F_U+F_\phi$ is evaluated the link $U(x,\mu)$ can be updated 
using SU(2) subgroups over-relaxation \cite{brown}.

\subsection{Even-odd preconditioning}

As we will see in Section 5.4,
the computational effort for obtaining decorrelated gauge configurations is proportional to $n^2$ (for fixed quark mass), where $n$ is the number of bosonic fields. The number of bosonic fields needed in the approximation depends on the 
condition number of the fermion matrix. A
way to lower the number of fields is to perform a preconditioning of the fermion matrix. To preserve the locality of the updating algorithm it is necessary that the preconditioned matrix remain local. A well-known possibility is even-odd preconditioning
. The lattice is labeled by even and odd points and a field $\phi$ can be written in a even-odd basis as
\be
\phi=\left(\begin{array}{c}
\phi_e\\\phi_o\end{array}\right)
\ee
where $\phi_e$ ($\phi_o$) denotes the part of the field $\phi$ living on the even (odd) sublattice, respectively. 
In the even-odd basis the fermion matrix $D$ can be written as
\be
D=\left(\begin{array}{cc}
1 & -KM_{eo}\\
-KM_{oe} & 1\end{array}\right)
\ee
where $M_{eo}$ and $M_{oe}$ are the off-diagonal part of the fermion matrix 
(\ref{fermionmatrix}) connecting odd to even sites or even to odd sites, respectively. Using the identity
\be
\det\left(\begin{array}{cc}
X & Y\\
W & Z\end{array}\right)
=\det X\det\left(Z-WX^{-1}Y\right)\label{trick}
\ee
(where $X,Y,W$ and $Z$ are matrices) we obtain
\be
\det D=\det\left(1-K^2M_{oe}M_{eo}\right)=:\det \hat D
\ee
The preconditioned matrix $\hat D$ is local and lives only on half of the lattice and involves next-to-nearest neighbor interactions. To simulate full QCD we could introduce a local bosonic action using the preconditioned fermion matrix $\hat D$.
\be
\hat S_b=\sum_{k=1}^n \phi_k^\dagger (\hat D-z_k)^\dagger(\hat D-z_k)\phi_k
\ee
but unfortunately the $\hat D^\dagger \hat D$ term connects $(\mbox{next})^3$-to-nearest neighbor sites, which makes the local updating too complex. A very simple way to save the situation is to use eq. (\ref{trick}) again 
and observe that  
\bea
&&\det (\hat D -z_k)=
\det\left((1-z_k)-K^2M_{oe}M_{eo}\right)=\nonumber\\
&&=\det 
\left(\begin{array}{cc}
1-r_k & -KM_{eo}\\
-KM_{oe} & 1-r_k\end{array}\right)
=\det(D-r_k)\label{qqqqq}
\eea
where the root $r_k$ is chosen such that 
\be
(1-r_k)^2=1-z_k\label{newroots}
\ee
The last term of (\ref{qqqqq}) allows us  
to define the local bosonic action for the preconditioned case 
without using explicitly the matrix $\hat D$ but simply redefining the roots of the polynomial from $z_k $ to $r_k$ of the non-preconditioned local action (\ref{locac}). 
\be
\hat S_b=\sum_{k=1}^n \phi_k^\dagger (D-r_k)^\dagger(D-r_k)\phi_k
\ee
This simple trick is only possible in the non-hermitian version of the algorithm. In the hermitian version a more complicated method has to be used 
\cite{beat2}. \\

The Metropolis test of the preconditioned case must be performed using the 
$\hat D$ matrix. One uses in the test the acceptance (\ref{32}) or (\ref{df})
with the preconditioned matrix
\be
W=[\hat D'P(\hat D')]^{-1}\hat DP(\hat D)
\ee
where $P(\hat D)$ is the polynomial with the original $z_k$ roots.

\section{Results for the exact local bosonic algorithm.}

Numerical simulations using the exact local bosonic algorithm in the non-noisy version described in the previous section have been performed on $4^4$, $4^3\times 16$  and
$8^4$ lattices for different $\beta$ and $k$ values. For the $4^4$ and $4^3\times 16$ lattices we concentrated on $\beta=0$ and $\beta=5$ and for the $8^4$ only on $\beta=0$ and $\beta=5.3$. 
We have performed the majority of the simulations reported upon here at 
$\beta=0$ for the following reasons:\\
i) at $\beta=0$ we know the critical hopping parameter $K_c$ ($K_c = 1/4$), and
the form of the spectrum of the fermion matrix (it is contained in a circle
centered at $(1,0)$ of radius $4 K$, and appears to be distributed uniformly
within that circle). So the optimal location of the polynomial roots $z_k$ is
easy in that case: we choose them on the circle $c = 1, f = 0$ 
in eq.(\ref{rootdef})\\
ii) at $\beta=0$ there is no danger of running into the deconfined phase 
before reaching $K = K_c$. This makes the comparison among several volumes
simpler.\\
iii) $\beta=0$ represents the hardest case for the inversion of the fermion
matrix, and is thus a good test of our algorithm.\\

We explored the acceptance of the Metropolis correction test and the number of iteration of the BiCGStab algorithm used in the test for inverting $DP(D)$.   
The study was performed by varying the degree $n$ of the polynomial and the hopping parameter $k$. The results are shown in Table 1 to Table 3. The trajectory length of all the runs is chosen to be 10. The stopping norm of the residual of the BiCGStab is
chosen to be $10^{-18}$.\\

In Tables 1 and 2 we present the results obtained from simulations without 
using even-odd preconditioning, on $4^4$ and $8^4$ lattices respectively. 
One observes that the acceptance increases quite rapidly with the degree of 
the polynomial, above some threshold. On the other hand, the number of 
iterations needed to invert $DP(D)$ remains very low for high enough acceptance.
These data show clearly that the overhead due to the Metropolis test remains 
negligible provided that the degree of the polynomial is tuned to have 
sufficient acceptance.\\

In Table 3 we present the results obtained from simulations using 
even-odd preconditioning. For comparison, the parameters of the simulations 
were chosen equal to some simulation parameters presented in Table 1. 
The effect of the preconditioning is 
to improve the approximation so that fewer fields are needed for the same 
acceptance. From the data we see that the preconditioning reduces the number 
of fields by at least a factor two. \\

Comparative examples of the plaquette values obtained using the exact local 
bosonic algorithm and the HMC algorithm are given in Table 4. For any choice 
of polynomial the plaquette remains consistent with the HMC data. 
However, the autocorrelation increases rapidly when the acceptance of the 
Metropolis test decreases. To optimize the efficiency of the algorithm one has 
to tune the degree of the polynomial to obtain a certain acceptance. 
The choice of the optimal acceptance is discussed in detail in the next section.\\

In Fig. 1 we show the gain in work due to the preconditioning as a function of the inverse quark mass (in the range of $\kappa$ values explored, $m_q$ is 
nearly proportional to $(1-K/K_c)$).
The simulation is done at $\beta=0$ on a lattice $4^3\times 16$ with the optimal polynomial. The work is in units of fermion matrix multiplications.  
Simulations with HMC were performed at the same parameters to compare the two 
algorithms. It turned out that the exact preconditioned local bosonic 
algorithm is as efficient as the HMC. Of cource, a more comprehensive comparison
is needed.\\

In Table 5 we present a comparative example between the non-hermitian and the hermitian version of the local bosonic action, without preconditioning. The simulations are done on a $4^4$ lattice at $\beta=5$ and $K=0.15$. 
The superiority of the non-hermitian exact version is evident.

\begin{table}[p]
\label{tab1}
\begin{center}
\begin{tabular}{|c|c|c|c|c|c|c|}\hline
lattice & $\beta$ & $K$  & $n$& acceptance & $I_{BiCGStab}$ \\\hline
$4^4$   &  0      & 0.215&    &            &        \\
        &         &      & 20 &   0.04(3)  &  53.0(9) \\
        &         &      & 24 &   0.20(5)  &  3.00(1) \\
        &         &      & 28 &   0.43(7)  &  2.00(1) \\
        &         &      & 36 &   0.92(4)  &  1.00(1) \\
        &         &      & 40 &   0.96(3)  &  1.00(1) \\
        &         &      & 44 &   0.94(4)  &  1.00(1) \\
        &         &      & 48 &   0.98(2)  &  1.00(1) \\\hline
$4^4$   &  0      & 0.230&    &            &        \\
        &         &      & 44 &   0.20(6)  &  3.00(1) \\
        &         &      & 46 &   0.31(7)  &  2.42(2) \\
        &         &      & 48 &   0.40(7)  &  2.20(3) \\
        &         &      & 50 &   0.60(9)  &  1.71(1) \\
        &         &      & 52 &   0.65(7)  &  1.54(3) \\
        &         &      & 54 &   0.67(10) &  1.42(2) \\
        &         &      & 56 &   0.73(4)  &  1.09(1) \\
        &         &      & 60 &   0.78(6)  &  1.00(1) \\
        &         &      & 70 &   0.90(4)  &  1.00(1) \\\hline
$4^4$   & 5       & 0.150&    &            &          \\
        &         &      & 8  &   0.03(1)  &  65(8) \\
        &         &      & 10  &  0.22(2)  &  3.03(2) \\
        &         &      & 12  &   0.46(3)  &  2.00(1) \\
        &         &      & 16  &   0.81(2)  &  1.00(1) \\
        &         &      & 20  &   0.93(1)  &  1.00(1) \\\hline
$4^4$   & 5       & 0.160&     &            &          \\
        &         &      & 12  &   0.09(2)  &  43(5) \\
        &         &      & 14  &   0.20(2)  &  3.10(4) \\
        &         &      & 16  &   0.45(3)  &  2.00(1) \\
        &         &      & 20  &   0.75(2)  &  1.00(1) \\
        &         &      & 24  &   0.86(2)  &  1.00(1) \\\hline
\end{tabular}
\end{center}
\caption{Acceptance and number of iterations of the BiCGStab algorithm $I_{BiCGStab}$. $4^4$ lattice simulation without even-odd preconditioning.}
\end{table}

\begin{table}[p]
\label{tab2}
\begin{center}
\begin{tabular}{|c|c|c|c|c|c|c|}\hline
lattice & $\beta$ & $K$  & $n$& acceptance & $I_{BiCGStab}$ \\\hline

$8^4$   &  0      & 0.215&    &            &        \\
        &         &      & 28 &   0.00(1)  &  75(1) \\
        &         &      & 32 &   0.15(8)  &  3.52(10) \\
        &         &      & 36 &   0.45(1)  &  2.00(2) \\
        &         &      & 40 &   0.65(8)  &  1.25(3) \\
        &         &      & 44 &   0.85(8)  &  1.00(1) \\
        &         &      & 48 &   0.93(4)  &  1.00(1) \\
        &         &      & 52 &   0.98(2)  &  1.00(1) \\
        &         &      & 56 &   0.97(3)  &  1.00(1) \\\hline
$8^4$   &  5.3    & 0.156&    &            &        \\
        &         &      & 28 &   0.42(1)  &  2.00(1) \\
        &         &      & 34 &   0.74(1)  &  1.00(1) \\
        &         &      & 40 &   0.87(1)  &  1.00(1) \\\hline
$8^4$   &  5.3    & 0.158&    &            &        \\
        &         &      & 24 &   0.04(1)  &  8.01(1) \\
        &         &      & 32 &   0.43(1)  &  1.97(3) \\
        &         &      & 38 &   0.71(1)  &  1.00(1) \\
        &         &      & 44 &   0.86(1)  &  1.00(1) \\\hline
\end{tabular}
\end{center}
\caption{Acceptance and number of iterations of the BiCGStab algorithm $I_{BiCGStab}$. $8^4$ lattice simulation without even-odd preconditioning.}
\end{table}

\begin{table}[p]
\label{tab3}
\begin{center}
\begin{tabular}{|c|c|c|c|c|c|c|}\hline
lattice & $\beta$ & $K$  & $n$& acceptance & $I_{BiCGStab}$ \\\hline

$4^4$   &  0      & 0.215&    &            &        \\
        &         &      & 6  &   0.00(1)  &  88(1) \\
        &         &      & 10 &   0.41(2)  &  3.24(2) \\
        &         &      & 14 &   0.74(1)  &  2.00(1) \\
        &         &      & 20 &   0.96(1)  &  1.00(1) \\\hline
$4^4$   &  0      & 0.230&    &            &        \\
        &         &      & 18 &   0.22(2)  &  4.75(3) \\
        &         &      & 24 &   0.60(1)  &  2.48(4) \\
        &         &      & 30 &   0.75(1)  &  1.87(2) \\
        &         &      & 36 &   0.88(1)  &  1.00(1) \\\hline
\end{tabular}
\end{center}
\caption{Acceptance and number of iterations of the BiCGStab algorithm $I_{BiCGStab}$. $4^4$ lattice simulation with even-odd preconditioning.}
\end{table}

\begin{table}[p]
\label{tab4}
\begin{center}
\begin{tabular}{|c|c|c|c|c|c|c|}\hline
lattice & $\beta$ & $K$  & $n$& plaq & $\tau_{int}$ \\\hline

$4^4$   &  5      & 0.15 &    &            &        \\
        &         &      & 8  &   0.413(10) &  58(5) \\
        &         &      & 10 &   0.416(2)  &  6(1) \\
        &         &      & 12 &   0.416(2)  &  5(1) \\
        &         &      & 16 &   0.415(2)  &  3(1) \\
        &         &      & 20 &   0.416(2)  &  3(1) \\
        &         &      &  &     &   \\
        &         &      & HMC &   0.4155(8)  &  7(1) \\
        &         &      &  &     &   \\\hline
$8^4$   &  5.3    & 0.158&    &            &        \\
        &         &      & 24 &   0.4933(?)  &  $>$300 \\
        &         &      & 32 &   0.4918(9)  &  40(5) \\
        &         &      & 38 &   0.4909(4)  &  18(2) \\
        &         &      & 44 &   0.4906(4)  &  16(2) \\
        &         &      &  &     &   \\
        &         &      & HMC &   0.4909(2)  &  ? \\
        &         &      &  &     &   \\\hline

\end{tabular}
\end{center}
\caption{Comparative examples between the plaquette values obtained using our 
exact local bosonic algorithm and the HMC algorithm. The integrated 
autocorrelation time is given in units of trajectories. Question marks denote 
informations which are 
not available or can not be determined. The acceptance of 
the Metropolis test for these simulations is given in Tables 1 and 2. 
The HMC data on the $8^4$ lattice is taken from [12]. }
\end{table}

\begin{table}[p]
\label{tab4+1}
\begin{center}
\begin{tabular}{|c|c|c|c|}\hline
algorithm     &   polynom& plaquette & acceptance \\\hline
non-hermitian & $n=8$                &   0.413(10) &  0.03(1)\\
exact         & $n=10$               &   0.416(2)  &  0.22(2)\\
version       & $n=12$               &   0.416(2)  &  0.46(3)\\
              & $n=16$               &   0.415(2)  &  0.81(2)\\
              & $n=20$               &   0.416(2)  &  0.93(1)\\\hline
hermitian     & $n=6,\epsilon=0.05$  &   0.4239(5) &  - \\
version       & $n=14,\epsilon=0.05$ &   0.4180(7) &  - \\
              & $n=50,\epsilon=0.05$ &   0.4162(12)&  - \\
              & $n=80,\epsilon=0.05$ &   0.4163(10)&  - \\
              & $n=50,\epsilon=0.005$&   0.413(1)  &  - \\
              & $n=80,\epsilon=0.005$&   0.413(2)  &  - \\
              & $n=50,\epsilon=0.01$ &   0.415(1)  &  - \\
              & $n=80,\epsilon=0.01$ &   0.415(1)  &  - \\\hline
              &                      &             &   \\
HMC           &                      &   0.4155(8) &    \\
              &                      &             &   \\\hline
\end{tabular}
\end{center}
\caption{ Comparative example between the non-hermitian exact local bosonic algorithm, the hermitian local bosonic algorithm and the HMC. Simulation on a $4^4$ lattice at $\beta=5$ and $K=0.15$. The autocorrelation time of the data in the hermitian case could not always be determined. For $n\geq 50$ it is of the order of some hunderts of sweeps. The autocorrelation of the data in the non-hermitian case is reported in the Table above.}
\end{table}

\section{Predicting the Metropolis acceptance}

Let us write generically the Metropolis acceptance probability (eq. (\ref{32}) or (\ref{df})) as
$P_{acc} = \min(1, e^{-\Delta E})$, with 
$\Delta E = \chi^\dagger (W^\dagger W - 1) \chi$. 
Detailed balance requires $< e^{-\Delta E} > = 1$,
so that $<(\Delta E)^2> \simeq 2 < \Delta E >$. Then, assuming the distribution of
$\Delta E$ to be Gaussian, which is certainly reasonable for systems large compared
to the correlation length $\sim m_q^{-1}$, one can easily derive (see Appendix A) 

\begin{equation}
< P_{acc} > \simeq erfc\left( \sqrt{\frac{<(\Delta E)^2>}{8}}  \right)
\label{accept}
\end{equation}

We can estimate $<(\Delta E)^2>$ here, because the Chebyshev-like approximation
we use for $D^{-1}$ has a known error bound. Recall (\cite{forcrand1}, eq.(29)) 
that for any eigenvalue $\lambda$ of $D$, one has

\begin{equation}
\alpha \equiv | \lambda P(\lambda) - 1 | \leq 
\tilde{\alpha} \equiv 
2~\left( \frac{a + \sqrt{a^2 - c^2}}{d + \sqrt{d^2 - c^2}} \right)^{n+1}
\label{error}
\end{equation}
where the spectrum of $D$ is contained in an ellipse ${\partial\cal S}$
of center $(d,0)$, 
large semi-axis $a < d$, and focal distance $c \le a$. For simplicity we
consider here the strong coupling case $\beta = 0$. Then the spectrum of
$D$ is contained in a circle ($c = 0$), and 
$\frac{a}{d} = \frac{K}{K_c}$, with $K_c(\beta=0) = 1/4$.
One can then expect, for the eigenvalues $\tilde{\lambda}$ of $(W^\dagger W - 1)$
which contains 4 factors of the form $D P(D)$:

\begin{equation}
| \tilde{\lambda} |  \leq 4 \tilde{\alpha}
\end{equation}
Then, assuming that the eigenvalues $\tilde{\lambda}$ fluctuate independently
of each other, it follows $<(\Delta E)^2> \leq 192 V \tilde{\alpha}^2$, where
$V$ is the number of lattice sites, or at $\beta = 0$:

\begin{equation}
<(\Delta E)^2> \leq 768 V \left(\frac{K}{K_c}\right)^{2 n + 2}
\end{equation}
In fact, the bound (\ref{error}) becomes much tighter for interior eigenvalues
of $D$, far from the boundary ${\partial\cal S}$. 
To model this effect, we introduce 
a ``fudge factor'' $f \leq {\cal O}(1)$ in $\tilde{\alpha}$. It will also
account for possible correlations between $D$ and $D'$ at the beginning
and end of a trajectory (ie. between successive Metropolis tests).
Substituting in (\ref{accept}) we finally obtain

\begin{equation}
< P_{acc} > \simeq erfc\left( f \sqrt{96 V} 
\left(\frac{K}{K_c}\right)^{n + 1} \right)
\label{fitacc}
\end{equation}

It becomes clear then that the acceptance will change very abruptly with the
number $n$ of bosonic fields, since it has the form $e^{-e^{-n}}$.

If our modeling of the Metropolis acceptance is correct, we should expect
$f$ to depend smoothly on $\beta$, but very little on $n, V$ and $K$.
Figs. 2, 3 and 4 show the acceptance we measured during our Monte-Carlo
simulations at $\beta = 0$, as a function of $n$, for 2 different volumes 
and 2 different $K$'s. The fit (eq. (\ref{fitacc})) is shown by the dotted
lines. All 3 figures have been obtained with the same value $f =0.19$.
We thus consider our ansatz (\ref{fitacc}) quite satisfactory.
At other values of $\beta$, one can fix $f$ by a test on a small lattice,
and then predict the acceptance for larger volumes and different quark masses.
We will use our ansatz below to analyze the cost of our algorithm.

\section{Understanding the dynamics}

        The coupled dynamics of the gauge and bosonic fields in L\"uscher's
method are rather subtle. The long autocorrelation times 
$\propto n / \sqrt{\epsilon}$ in the Hermitian formulation were only explained
a posteriori \cite{borrelli}. In this section we follow a similar approach for 
our non-Hermitian formulation: we report our numerical observations, then
try to explain them using a simple toy model.

\subsection{Autocorrelation times}

        The Metropolis test, while ensuring that our algorithm is exact,
does not affect the dynamics of the gauge and bosonic fields, and obscures
their analysis. Therefore we remove this step from the simulations reported
in this section. We measure the integrated autocorrelation time $\tau_{int}$
of the plaquette, in units of sweeps.
We tried to maintain, in each case,
a total number of sweeps of at least $100 \tau_{int}$; nonetheless we still
expect sizeable 
errors in our $\tau_{int}$ estimates (more than $ 10\%$).\\

        We first varied the number $n$ of bosonic fields, while keeping
$\beta = 0$ and $K = 0.215$. Our data is shown by $\times$ and $+$
in Fig. 5, for a $4^4$ and $4^3 \times 16$ lattice respectively.
Independently of the volume, we get $\tau_{int} \sim 1.3 n$. On the other
hand, if we freeze the bosonic fields, updates of the gauge fields 
decorrelate the plaquette in ${\cal O}(1)$ sweep. The situation
thus appears identical to the Hermitian case. Our explanation
then \cite{borrelli} was that the slow dynamics were entirely 
caused by the bosonic
fields, and that the autocorrelation time for the whole system was simply
$\max_k \xi_k^z$, where $\xi_k$ is the $\phi_k$ correlation length and
$z \sim 2$ the dynamic critical exponent. This scenario explained all 
our data. Here we test that scenario by applying a global heatbath at 
each sweep to each $\phi_k$. Namely we update the bosonic fields by
the assignment $\phi_k \longleftarrow (D - z_k)^{-1} \chi_k$, where
$\chi_k$ is an independent Gaussian vector. While costly, this kind of
updating should reduce the autocorrelation time of the bosonic dynamics,
and thus of the whole system, to ${\cal O}(1)$ sweep. To our surprise,
our measurements, shown by $\star$'s in Fig. 5, indicate that $\tau_{int}$
still grows linearly with $n$, and is only reduced by a factor $\sim 2$
compared to a local update. This is clearly caused by the coupling of
all bosonic fields to each other, through the gauge field.

\subsection{Toy model}

        To understand this phenomenon better, we consider a toy model
of $n$ scalar complex variables $\tilde{\phi_k}$, and one scalar (complex
or real) variable $x$, with action

\begin{equation}
S_{toy} = \sum_{k=1}^n | (x - z_k) \tilde{\phi_k} |^2
\end{equation}

The single variable $x$ thus plays the role of the Dirac operator $D$,
and represents all the gauge degrees of freedom, while the $\tilde{\phi_k}$'s
represent all the bosonic fields. The zeroes $\{z_k\}$ are the same as 
in eq.(\ref{locac}).

Consider Monte Carlo updates of this toy model, where the $\tilde{\phi_k}$'s
are updated by a heatbath, and $x$ by over-relaxation. Namely, at each sweep:

\be
\tilde{\phi_k} \longleftarrow \frac{1}{x - z_k} \tilde{\chi_k}
\label{update_toy1}
\ee
\be
x \longleftarrow 
\frac{2 \sum_k z_k |\tilde{\phi_k}|^2}{\sum_k |\tilde{\phi_k}|^2} - x
\label{update_toy2}
\ee

Substituting $(\ref{update_toy1})$ into $(\ref{update_toy2})$ gives
\begin{equation}
x_{new} - x = -2 \frac{\sum_k \frac{1}{\bar x - \bar z_k} |\tilde{\chi_k}|^2}
{\sum_k \frac{1}{|x - z_k|^2} |\tilde{\chi_k}|^2}
\end{equation}

Since $|\tilde{\chi_k}|^2 \sim 1$, one notices that the numerator
above is $\sim -1 / x$. On the other hand, since $|x - z_k|$ is bounded
from above ($x$ must stay inside ${\partial\cal S}$), the 
denominator is bounded from {\em below} by some constant times $n$.
The step size of the $x$ updates will therefore decrease like $1/n$,
and the autocorrelation time increase like $n$, in spite of the heatbath
on the $\tilde{\phi_k}$'s.\\

This analysis carries through whether $x$ is complex or real. There is
a difference however, in the minimum distance $min_k |x - z_k|$. 
If $x$ is complex, it will stay most of the time well
inside the domain ${\partial\cal S}$, and will only approach one of the
zeroes $z_k$ occasionally. But if $x$ is real, which corresponds to the
Hermitian algorithm, $x$ will always be near one of the zeroes $z_k$, 
because they all lie a distance $\sim 1/\sqrt{\epsilon}$ from the real axis,
at intervals $\sim 1/n$.  Therefore we should expect an additional slowing
down $\sim 1/\sqrt{\epsilon}$ in this case, whereas the choice of ``$K$''
has little effect in the non-Hermitian case. These expectations are fully
confirmed by numerical simulations of our toy model.\\

Going back to our original QCD action, we now understand why a global
heatbath of the bosonic fields fails to decorrelate the gauge fields.
We also expect another advantage of the non-Hermitian formulation:
only a few modes of the gauge field, those with eigenvalues near 
${\partial\cal S}$, will suffer additional slowing down (beyond a factor $n$)
as $K$ approaches $K_c$; in the Hermitian formulation, {\em all}
modes will slow down as $n / \sqrt{\epsilon}$. Unfortunately, we have not found
any way to accelerate the dynamics in either case, even in our toy model.

\subsection{Cost of the simulation}

Since we now have a reasonable understanding of the Metropolis acceptance and
of the dynamics without the Metropolis test, we can see the effect of the one
on the other, and then estimate the total cost of the algorithm per independent
configuration.

Denote by $W_i$ the value of our observable (the plaquette) at the beginning
of trajectory $i$, and by $W_i'$ the value for the candidate configuration
proposed to the Metropolis test at the end of trajectory $i$. Then the 
probability distribution for $W_{i+1}$ is

\begin{equation}
P(W_{i+1}) = P_{acc} \delta(W_i') + (1 - P_{acc}) \delta(W_i)
\end{equation}
Assume for simplicity that $<W> = <W'> = 0$. The autocorrelation of $W$ will
then be

\begin{equation}
\frac{<W_{i+1} W_i>}{<W>^2} = <P_{acc}> \frac{<W_i' W_i>}{<W>^2} + (1 - <P_{acc}>)
\end{equation}
where we neglected correlations between $W_i'$ and $P_{acc}$. Calling $\tau$
and $\tau_0$ the autocorrelation time with and without Metropolis respectively,
we obtain

\begin{equation}
e^{-1/\tau} = <P_{acc}> e^{-m/\tau_0} + 1 - <P_{acc}>
\end{equation}
or

\begin{equation}
\tau = \frac{-1}{\log(1 - <P_{acc}> (1 - e^{-m/\tau_0}))}
\label{tau}
\end{equation}
for a trajectory of $m$ sweeps. Note that $\tau_0$ is measured in sweeps and 
$\tau$ in trajectories. Folding into (\ref{tau}) our ansatz for $<P_{acc}>$ (eq. (\ref{fitacc})),
and using our empirical result $\tau_0 \sim 1.3 n$, we obtain in Fig. 4 
the autocorrelation time as a function of n, for lattices of sizes 4, 8, 16, 32, at
$\beta = 0$ and $K = 0.215$, with $m = 10$. The behavior will be qualitatively
similar for other choices of parameters. The Monte Carlo data shown in Fig. 6 was
obtained on a $4^4$ lattice, and is roughly compatible with eq.(\ref{tau}).\\

The operation count of the algorithm is about $6 n$ matrix-vector multiplications
by the Dirac operator $D$ per sweep. Therefore we can estimate the total cost of
our algorithm to produce an independent configuration, measured in multiplications
by $D$ per lattice site, as a function of the number of fields or the Metropolis
acceptance. This is shown in Figs. 7 and 8 respectively, for lattices of size
$L = 4$ to $32$. Fig. 7 shows, as expected, that the optimal number of bosonic
fields grows logarithmically with the volume.  The minimum work per site varies
by a factor $< 4$ going from $L = 4$ to $32$. Recall that in HMC, the work per
site is expected to grow like $L$, so it would increase by a factor 8 under the
same change of volume. The asymptotic advantage of our algorithm thus translates
into
a slim factor $\sim 2$ for reasonable lattice sizes.  Fig. 8 gives the useful
hint that the acceptance should be kept high, $70 - 80\%$, but that the optimum
is fairly broad.  This high acceptance justifies a posteriori our neglect of
the overhead due to the linear solver in the Metropolis test, since the number
of iterations necessary is then ${\cal O}(1)$ (see Tables 1 and 2). 
It is also clear from this Figure that the Metropolis test actually {\em saves} a lot
of work: if one wanted to get quasi-exact results (ie. acceptance $\sim 1$) without it,
one would pay a factor $>_{\sim} 4$ penalty, caused by the necessary increase both
in the number of bosonic fields and in the autocorrelation time.
Finally, it is instructive to compute the error bound (\ref{error}) at the optimal 
acceptance point, since it is the analogue of the parameter $\delta$ of the original 
L\"uscher proposal \cite{luescher1}: we get $\sim 1.4\%$ or $0.01\%$, for lattices of 
size 4 or 16; when the error becomes that small, it becomes most efficient 
to make the algorithm exact.

\subsection{Scaling}

The scaling of the number of bosonic fields and of the total complexity with
the volume and the quark mass has already been discussed in \cite{forcrand2}.
Our analysis confirms these earlier estimates.\\

$\bullet$ As the volume $V$ increases, the number $n$ of bosonic fields should
grow like $\log V$. This is a consequence of the exponential convergence of the
polynomial approximation used (eq.(\ref{error})), and is clearly necessary to 
keep a constant Metropolis acceptance (eq.(\ref{fitacc})). Since the work per
sweep is proportional to $n V$, and the autocorrelation time to $n$, the work
per independent configuration grows like $V (\log V)^2$.
This is a slower growth than Hybrid Monte Carlo which requires work $\sim V^{5/4}$.
But this is more of an academic than a practical advantage, as discussed above.\\

$\bullet$ As the quark mass $m_q$ decreases, the number of fields $n$ must grow like
$m_q^{-1}$. This can be derived from eq.(\ref{error}) or (\ref{fitacc}), using
$m_q \propto 1 - K/K_c$, which applies for small quark masses. 
The work per sweep is proportional to $n$. The autocorrelation time 
behavior is less clear. We expect that the autocorrelation behaves like
$\tau \sim n m_q^{-\alpha}$, since $m_q$ enters in the effective mass term
of each harmonic piece of the action $S_b$ (eq. \ref{locac}), and since
a factor $n$ comes from the 
autocorrelation of the gauge fields alone (remember the toy model in 
section 5.2).   
Our toy model does not say
anything about local $\phi$ updates, so to determine $\alpha$ our main 
argument comes from MC data. From an exploratory simulation on a $4^4$ 
lattice at $\beta=0$ (see Table 6) we obtained 
that $\alpha$ is near $1$. This is only 
indicative, because the dependence of $\alpha$ with the volume and the 
coupling has yet to be explored. \\

\begin{table}
\label{tab5}
\begin{center}
\begin{tabular}{|c|c|c|c|c|}\hline
K & $(1-K/K_c)$ & n &$\tau_{int}$& Work \\\hline
0.215 & 0.14 & 14 & 5 & 4200 \\
0.230 & 0.08 & 36 & 10 & 21600 \\\hline
\end{tabular}
\end{center}
\caption{ Exploratory measurement of the integrated autocorrelation time 
$\tau_{int}$ of the plaquette at $\beta=0$ on a $4^4$ lattice. 
The autocorrelation time is in units of trajectory. 
A trajectory is made of 10 sweeps. 
The degree of the polynomial is chosen to minimize the work. 
Each measurement is taken from a sample of 200000 sweeps. 
}
\end{table}

Another interesting exploratory comparison was made between local and  global heat-bath (see Table 7) without Metropolis test at $\beta=0$ on a $4^3\times 16$ lattice with fixed number of fields.
 We have measured the integrated autocorrelation of the plaquette 
for two quark masses. From the data in Table 7   
it is clear that the exponent $\alpha$ is much lower for the global heat-bath, as expected from global algorithms. Of cource, the cost of a global heath-bath is prohibitive, because the matrix $(D-z_k)$ has to be inverted for each boson field $\phi_k$\footnote{Its 
cost is at least proportional to $n/m_q$, and the prefactor turns out to be 
very large. This is due to the fact that the BiCGStab algorithm does not converge for all $D-z_k$, so that CG has to be used. Preconditioning this matrix using the polynomial $P(D)$
 will not improve the situation because all the cost is shifted in the 
preconditioning matrix. Only the prefactor can be reduced.}. 
These simulations indicate that accelerating the boson fields using a cheaper 
algorithm than the global heat-bath could improve the dynamics substantially.

\begin{table}
\label{tab6}
\begin{center}
\begin{tabular}{|c|c|c|c|c|}\hline
K & $(1-K/K_c)$ & n & \multicolumn{2}{c|}{$\tau_{int}$}\\ \cline{4-5}
  &  &  & local HB+OR & global HB\\ \hline
0.215 & 0.14 & 20 & 26 & 16 \\
0.230 & 0.08 & 20 & 65 & 22 \\
0.240 & 0.04 & 20 & 150 & ? \\ \hline
\end{tabular}
\end{center}
\caption{ Exploratory measurement of the integrated autocorrelation time 
$\tau_{int}$ of the plaquette at $\beta=0$ on a $4^3\times 16$ lattice, 
for different quark masses and for local and global update of the boson fields. 
The degree of the polynomial is the same for all runs. The simulation is 
performed without Metropolis test. 
The autocorrelation time is in units of sweeps.}
\end{table}

\newpage
\section{Conclusion}

We have presented an alternative algorithm to HMC for simulating dynamical 
quarks in lattice QCD. This algorithm is based on a local bosonic action. 
A non-hermitian polynomial approximation of the inverse of the quark matrix 
is used to define the local bosonic action. The addition of a global 
Metropolis test corrects the systematic errors. The overhead of the correction 
test is minimal. Even-odd preconditioning can also be used improving 
the approximation by at least a factor of two. Its implementation is trivial.\\

This algorithm is exact for any choice of the polynomial approximation. 
No critical tuning of the approximation parameters is needed. Only the efficiency
of the algorithm, which can be monitored, will be affected by the choice of parameters.
The cost of the algorithm increases with the volume $V$ of the lattice as $V(\log V)^2$
and with the inverse of the quark mass $m_q$ as $\frac{n^2}{m_q^\alpha}\sim 
\frac{1}{m_q^{2+\alpha}}$ with  
$\alpha\simeq 1$. 
This compares favorably with the scaling of HMC. \\

Further extensions of this algorithm can be explored. 
In particular, non-local preconditioning techniques can cheaply be used 
because the inverse of the preconditioning matrix is not needed, contrary
to HMC. A second possibility is to accelerate the dynamics of the boson fields,
using non-local algorithms cheaper than a global heat-bath 
(for example multi-grid). A direct acceleration of the gauge fields is 
still an open problem.
\newpage

{\Large {\bf Acknowledgments}}\\

We would like to thank the group of M. L\"uscher for discussions and for 
allowing access to part of their documentation. 
We thank F.Jegerlehner for his help during the first stage of the project and 
B. Jegerlehner for helpfull discussions. Finally, we thank A. Bori\c{c}i and S. Solomon for useful suggestions and the SIC of the EPFL together with the
support team of the Cray-T3D.\\[2cm]

{\Large {\bf Appendix: Proof of eq.(\ref{accept})}}\\

Detailed balance requires $< e^{-\Delta E} > = 1$,
so that $<(\Delta E)^2> \simeq 2 < \Delta E >$. We assume that the distribution of $\Delta E$ is Gaussian.
Let $x=\Delta E$ be a Gaussian distributed variable with variance $\sigma^2\simeq 2\bar x$. Its distribution is
$$
P(x)=\frac{1}{\sigma \sqrt{2}\pi}
\exp\left(-\frac{(x-\bar x)^2}{2\sigma^2}\right)
$$
From the Metropolis acceptance probability $P_{acc}=\min(1,e^{-\Delta E})$ we obtain
\bea
<P_{acc}>&&=\int_{-\infty}^0 dxP(x)+\int_0^\infty dxP(x)e^{-x}=\nonumber\\
&&=\frac{1}{\sigma\sqrt{2}\pi}\left[
\int_{-\infty}^0 dx e^{-\frac{(x-\bar x)^2}{2\sigma^2}}+
\int_0^\infty dx e^{-\frac{x^2-2x(\bar x-\sigma^2)+\bar x^2}{2\sigma^2}}\right]=\nonumber\\
&&=\frac{1}{\sigma\sqrt{2}\pi}\left[
\int_{-\infty}^{-\bar x} dy e^{\frac{-y^2}{2\sigma^2}}+
\int_{-\bar x+\sigma^2}^{\infty} dy e^{\frac{-y^2}{2\sigma^2}}
e^{-\bar x+\sigma^2/2}\right]
\nonumber
\eea
Using the fact that $-\bar x+\sigma^2\simeq \bar x$ and  $-\bar x+\sigma^2/2\simeq 0$ we obtain
\bea
<P_{acc}>\simeq &&\sqrt{\frac{2}{\pi}}\frac{1}{\sigma}\int_{\bar x}^\infty dy  
e^{\frac{-y^2}{2\sigma^2}}=\nonumber\\
&&
=\frac{2}{\sqrt{\pi}}\int_{\frac{\bar x}{\sigma\sqrt{2}}}^\infty dt e^{-t^2}=
erfc\left(\frac{\bar x}{\sigma\sqrt{2}}\right)=
erfc\left(\sqrt{\frac{<(\Delta E)^2>}{8}}\right)\nonumber
\eea

\begin{figure}
\centerline{\epsfysize = 8 in \epsffile {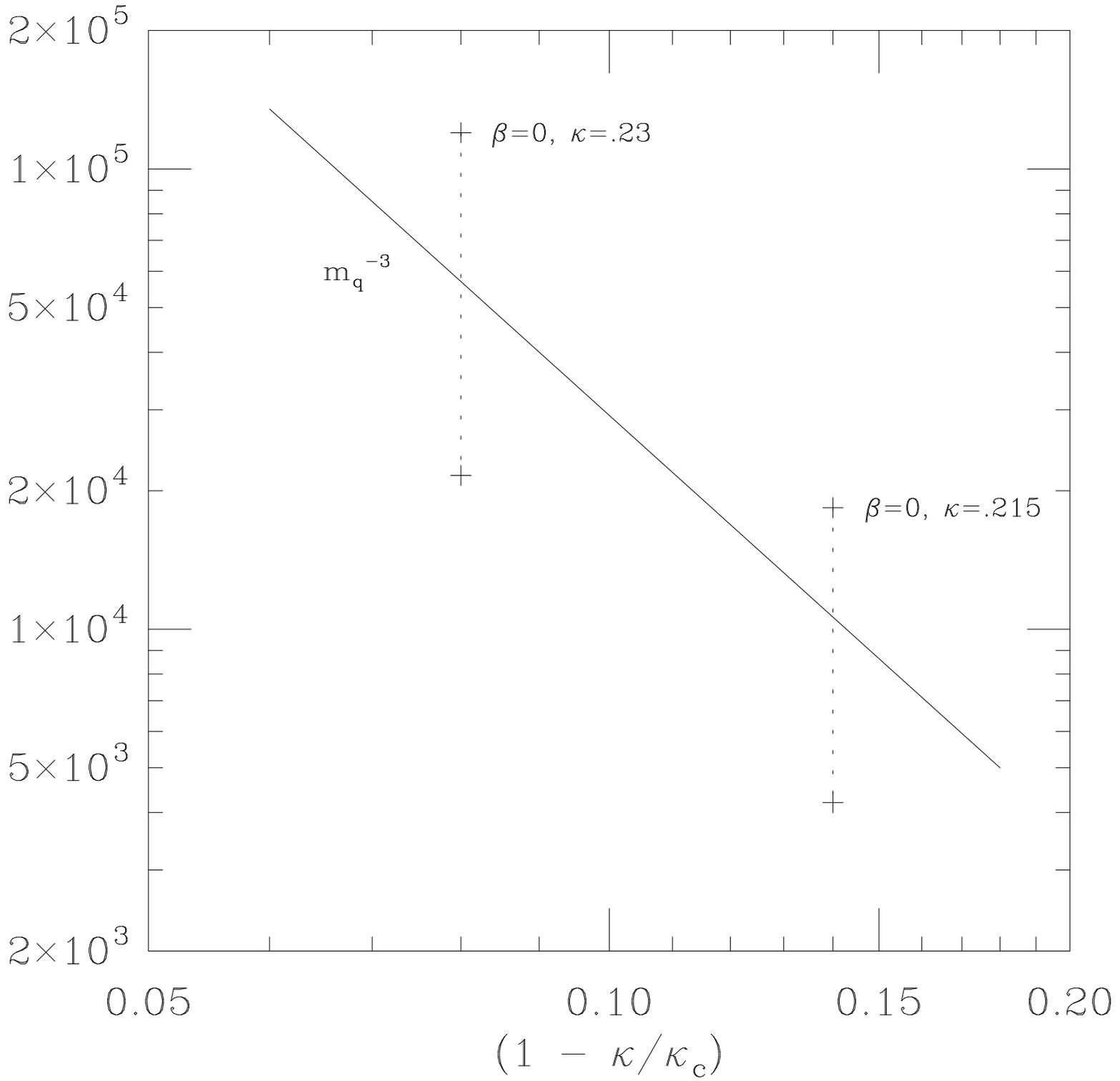}}
\caption{ Work per independent configuration as $K$ is varied.
Dotted lines indicate the gain due to even-odd preconditioning. The abscissa
is proportional to $m_q$ in the range of $K$ chosen. The work is
measured in applications of the Dirac operator per site. The lattice size
is $4^4$.}
\end{figure}

\begin{figure}
\centerline{\epsfysize = 8 in \epsffile {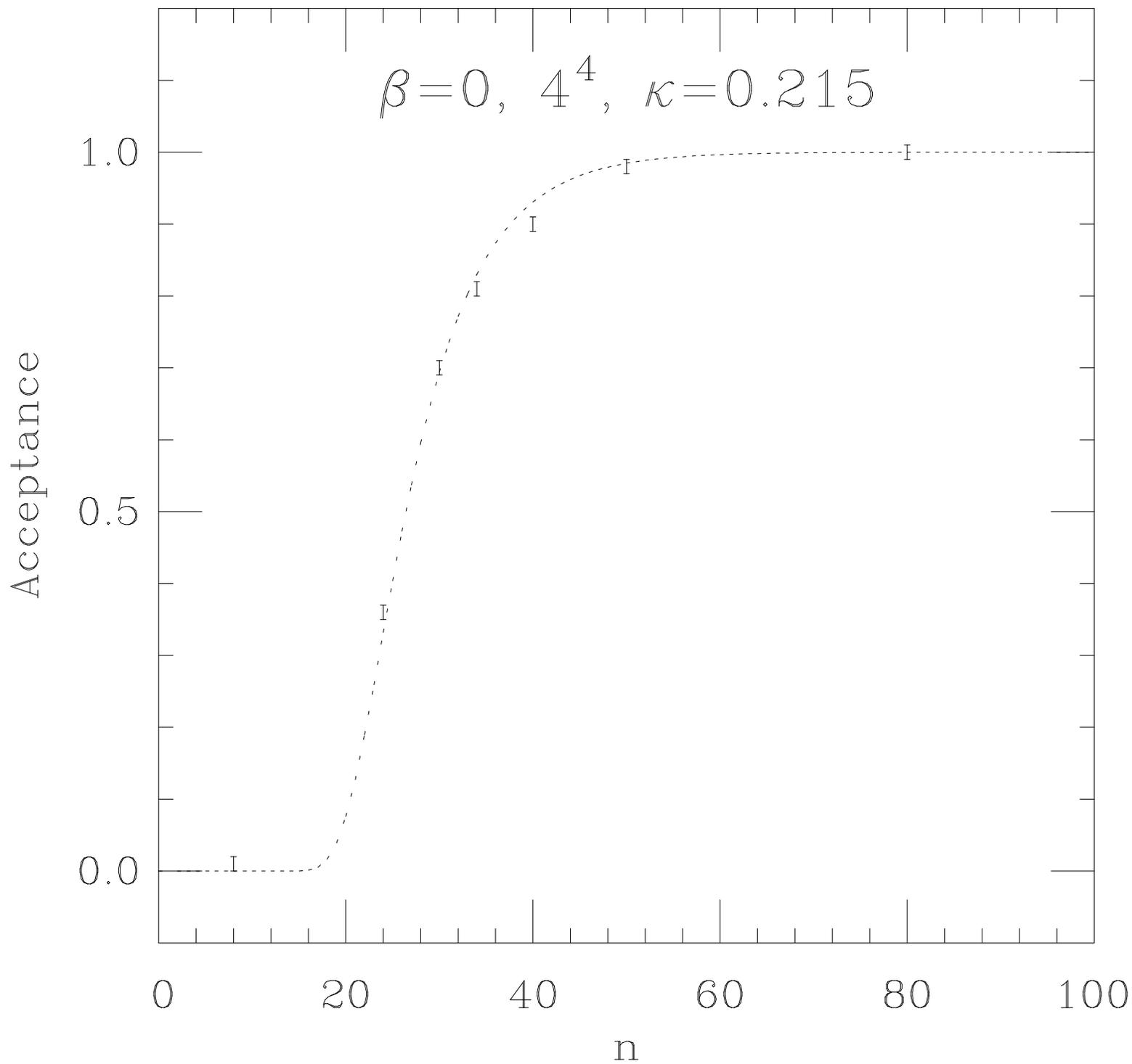}}
\caption{ Metropolis acceptance as a function of the number of bosonic
fields. The dotted line is our 1-parameter ansatz eq.(49).}
\end{figure}

\begin{figure}
\centerline{\epsfysize = 8 in \epsffile {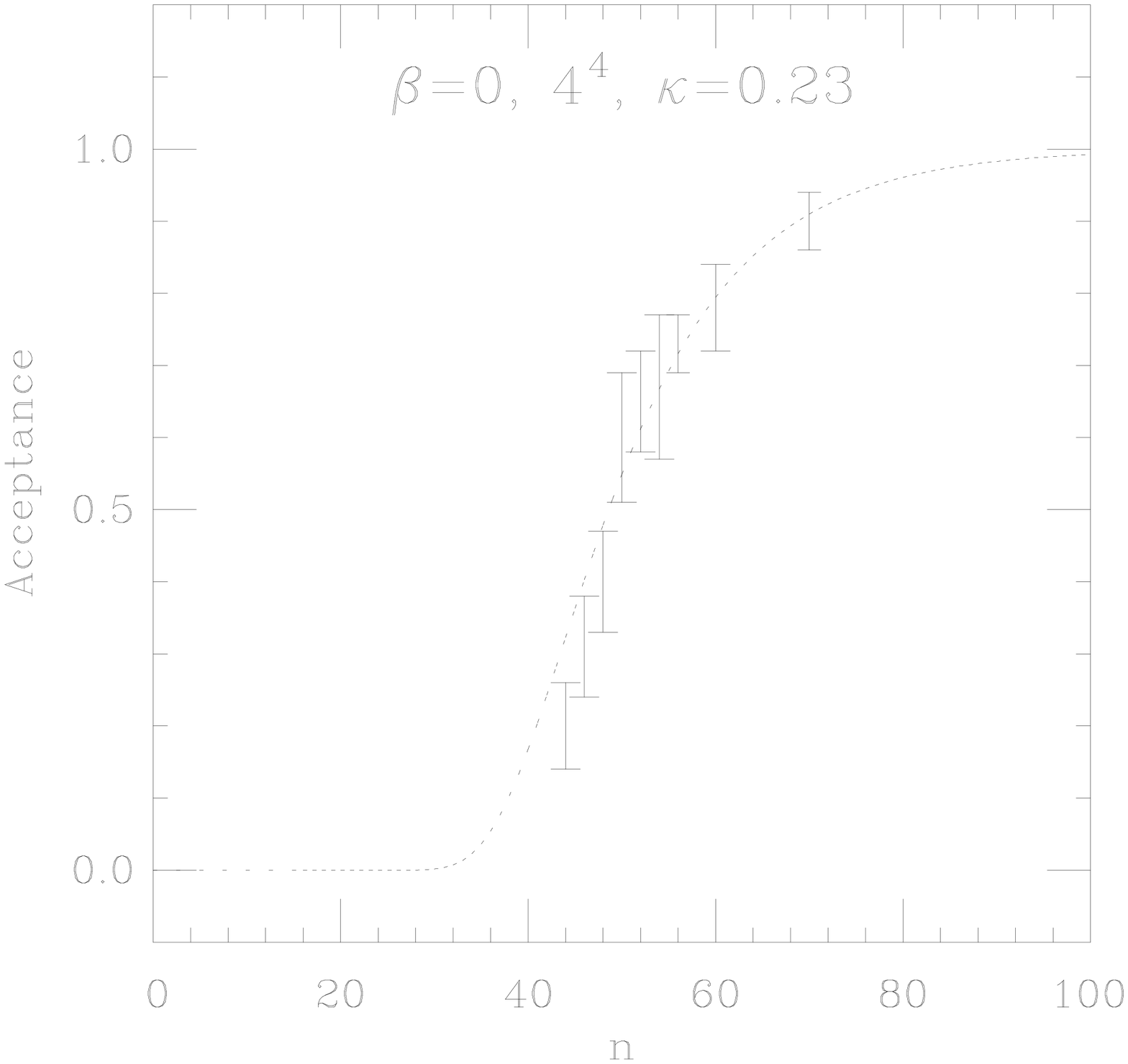}}
\caption{ Same as Fig.2, for a different value of $K$. The fit
parameter $f$ (eq.(49)) is unchanged from Fig.2.}
\end{figure}

\begin{figure}
\centerline{\epsfysize = 8 in \epsffile {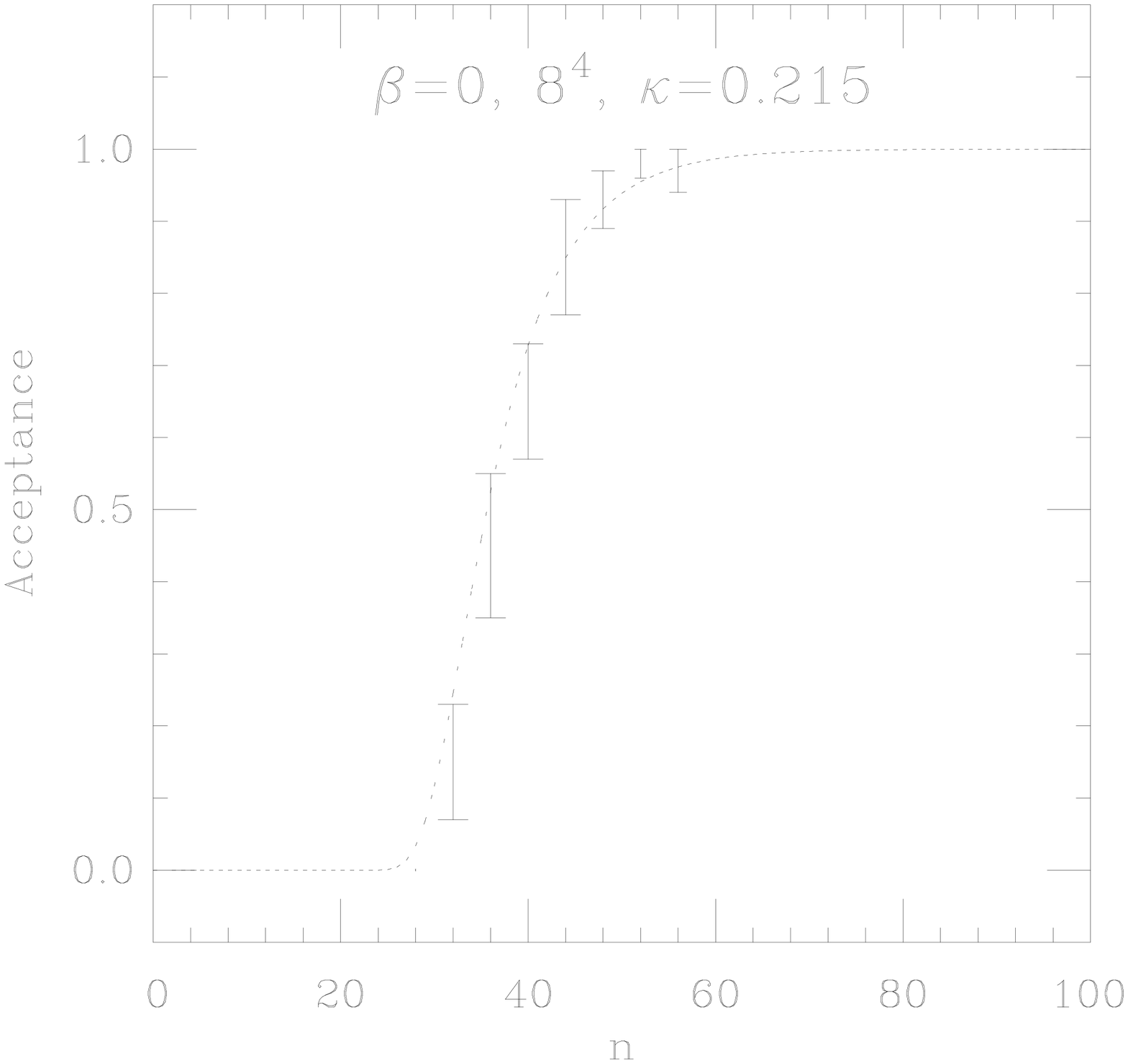}}
\caption{ Same as Fig.2, for a different volume. The fit parameter $f$
(eq.(49)) is unchanged from Fig.2.}
\end{figure}

\begin{figure}
\centerline{\epsfysize = 7.85 in \epsffile {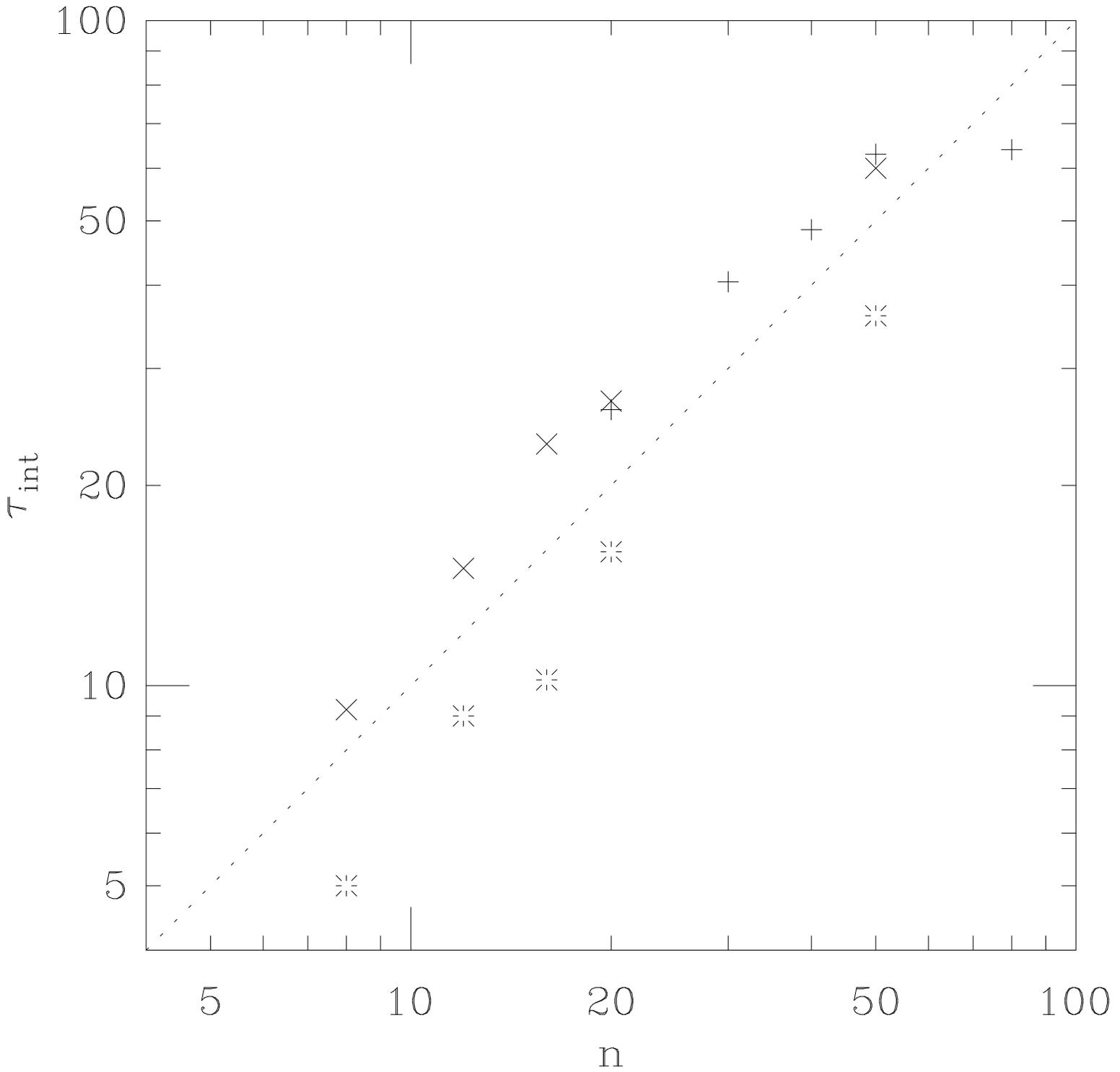}}
\caption{Integrated autocorrelation time of the plaquette, measured
in sweeps, versus the number of bosonic fields. The dotted line shows
$\tau_{int} = n$ to guid the eye. Crosses stand for a $4^4$ lattice, plusses
for a $4^3\times 16$ lattice, both under a {\em local} update of the bosonic fields.
Stars stand for a $4^3\times16$ lattice, under a {\em global} heatbath of
the bosonic fields. 
Simulations at $\beta = 0, K = 0.215$ without Metropolis test.}
\end{figure}

\begin{figure}
\centerline{\epsfysize = 8 in \epsffile {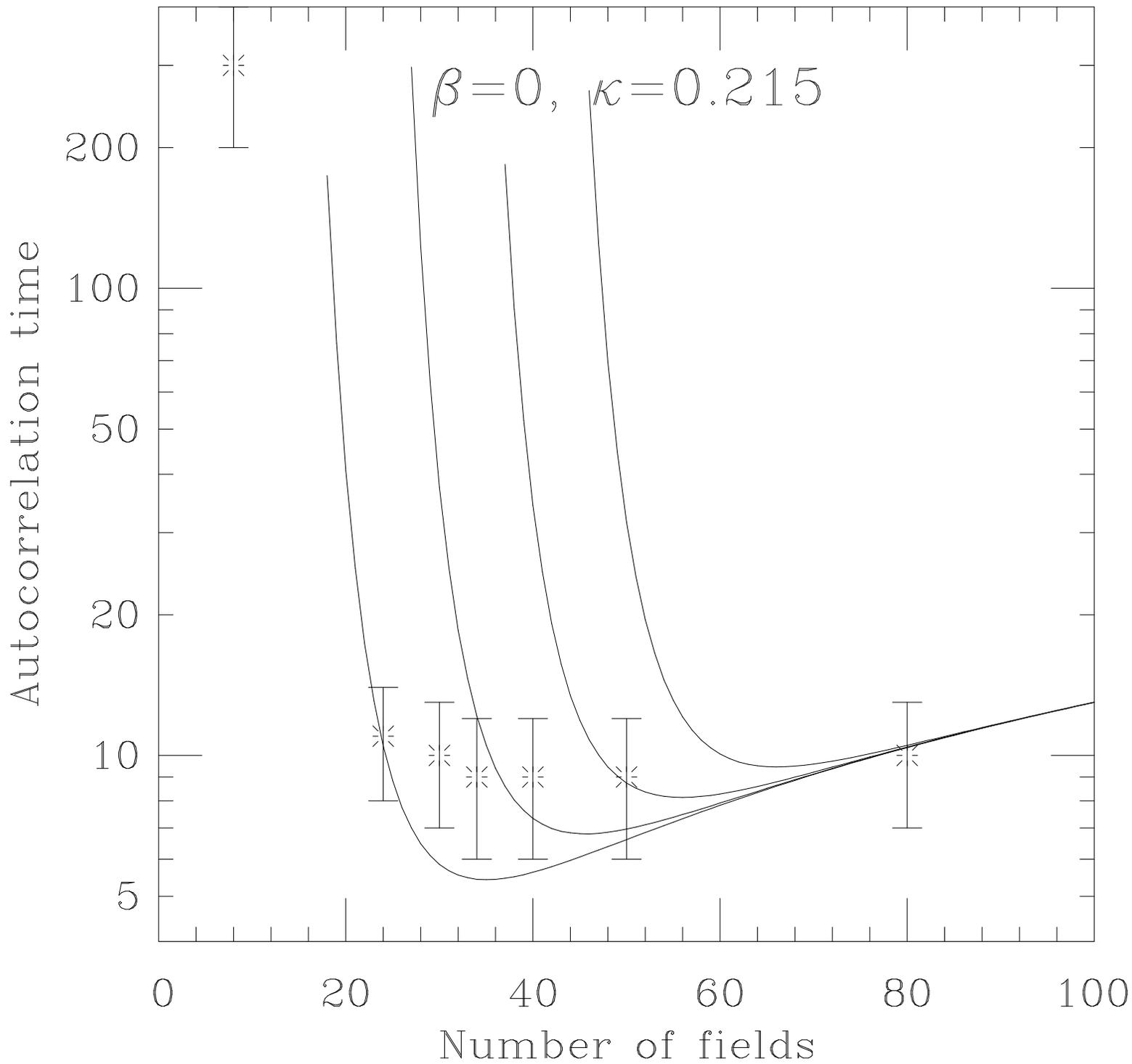}}
\caption{Integrated autocorrelation time given by our ansatz (58)
as a function of the number of fields, measured in trajectories, for lattices
of size $L = 4, 8, 16, 32$ from left to right. The Monte Carlo results also
shown have been obtained for $L = 4$.}
\end{figure}

\begin{figure}
\centerline{\epsfysize = 8 in \epsffile {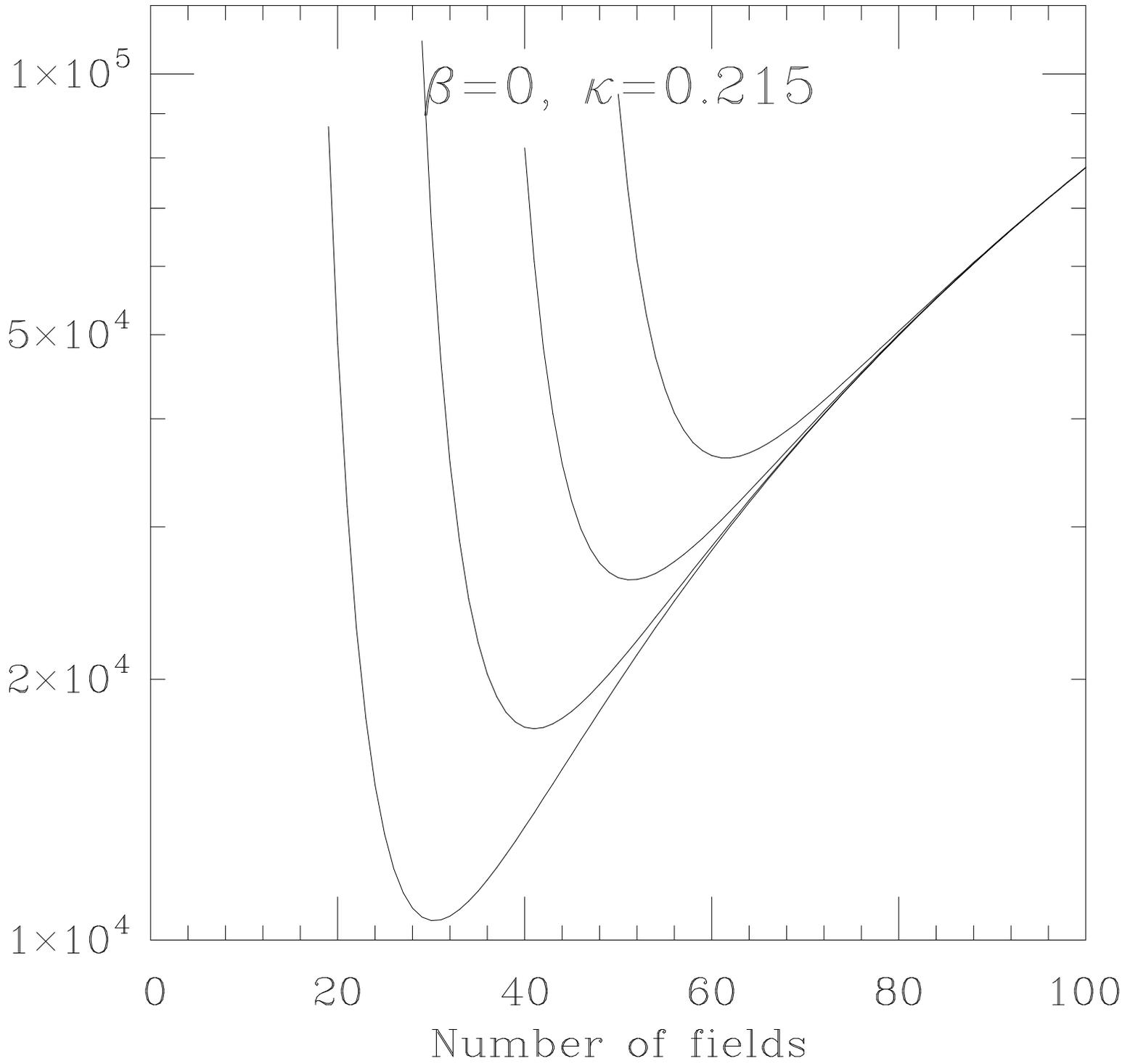}}
\caption{ Work per independent configuration versus the number of bosonic
fields, for lattices of size $L = 4, 8, 16, 32$ from left to right. The work
is measured in applications of the Dirac operator $D$ per site.}
\end{figure}

\begin{figure}
\centerline{\epsfysize = 8 in \epsffile {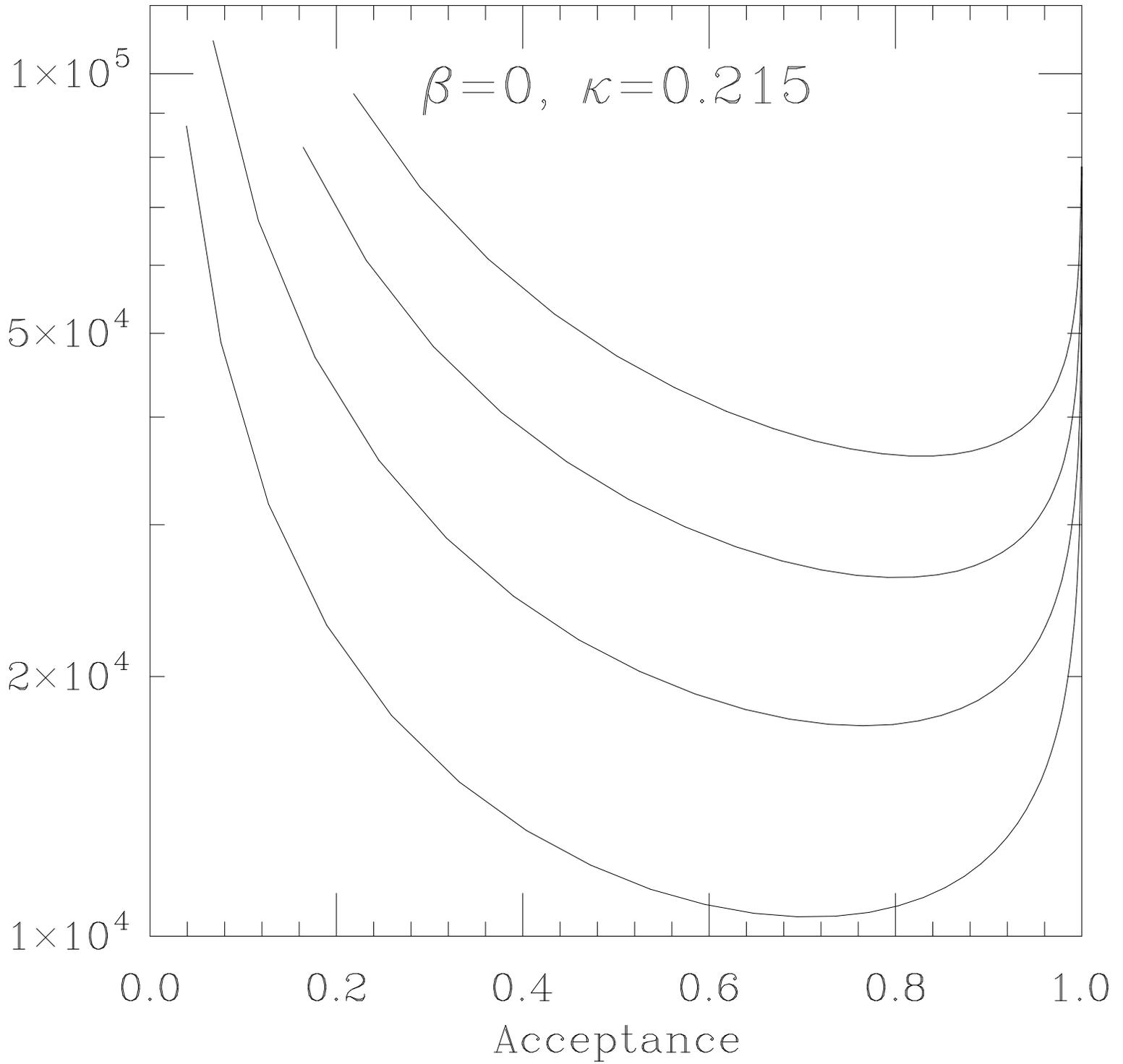}}
\caption{Same as Fig. 7, as a function of the Metropolis acceptance.}
\end{figure}

\end{document}